\documentclass[a4paper,11pt]{article}%
\usepackage{graphicx}
\usepackage{amsbsy}
\usepackage{amsmath}
\usepackage{cite}
\usepackage{slashed}
\usepackage{epsfig}%
\usepackage{amsfonts}%
\usepackage{amssymb}
\usepackage[colorlinks=true
,urlcolor=blue
,anchorcolor=blue
,citecolor=blue
,filecolor=blue
,linkcolor=blue
,menucolor=blue
,pagecolor=blue
,linktocpage=true
]{hyperref}

\begin{document}
\thispagestyle{empty}  \setcounter{page}{0}  
\begin{flushright}
LPSC14058\\
April 2014\\
\end{flushright}

\vskip   2.1 true cm

\begin{center}
{\huge Baryonic R-parity violation and its running}\\[1.9cm]

\textsc{J\'er\'emy Bernon}$^{1}$ \textsc{and Christopher Smith}$^{2}$%
\vspace{0.5cm}\\[9pt]\smallskip{\small \textsl{\textit{Laboratoire de Physique
Subatomique et de Cosmologie, }}}\linebreak
{\small \textsl{\textit{Universit\'{e} Grenoble-Alpes, CNRS/IN2P3, 53 avenue
des Martyrs, 38026 Grenoble Cedex, France}.}} \\[1.9cm]\textbf{Abstract}\smallskip
\end{center}

\begin{quote}
\noindent 
Baryonic R-parity violation arises naturally once Minimal Flavor Violation (MFV) is imposed on the supersymmetric flavor sector at the low scale. At the same time, the yet unknown flavor dynamics behind MFV could take place at a very high scale. In this paper, we analyze the renormalization group (RG) evolution of this scenario. We find that low-scale MFV is systematically reinforced through the evolution thanks to the presence of infrared fixed points. Intriguingly, we also find that if holomorphy is imposed on MFV at some scale, it is preserved by the RG evolution. Furthermore, low-scale holomorphy is a powerful infrared attractor for a large class of non-holomorphic scenarios. Therefore, supersymmetry with minimally flavor violating baryon number violation at the low scale, especially in the holomorphic case but not only, is viable and resilient under the RG evolution, and should constitute a leading contender for the physics beyond the Standard Model waiting to be discovered at the LHC.

\let   \thefootnote    \relax
\footnotetext{$^{1}\;$bernon@lpsc.in2p3.fr} 
\footnotetext{$^{2}\;$chsmith@lpsc.in2p3.fr}
\end{quote}

\newpage

\section{Introduction}

Supersymmetric particles have not yet shown up at the LHC. Though the current bounds on the masses of the supersymmetric particles depend on the assumed spectrum, they are now generally at or above the TeV scale. Such a large splitting in mass between Standard Model (SM) particles and their superpartners renders the model less radiatively stable, and requires delicate fine-tunings of its parameters to be viable. There is however one scenario in which these bounds are trivially relaxed. Most of the current searches assume that the minimal supersymmetric Standard Model (MSSM) incorporates R-parity as an exact global symmetry~\cite{FarrarF78}. As is well-known, see \textit{e.g.}~\cite{Barbier04} for a review, this forces superparticles to be present in pairs in all vertices, and thus renders the lightest superparticle (LSP) perfectly stable. Supersymmetric events are accompanied by a significant amount of missing energy carried away by the LSP. On the other hand, if R-parity is not exact, the LSP decays and these missing energy signatures are simply not there.

Removing R-parity may be fine for the LHC, but immediately allows for proton decay or neutron oscillations. The tight bounds on these observables imply that R-parity violating (RPV) couplings involving light fermion flavors must be really tiny. So, either the RPV couplings are globally suppressed, or they are highly hierarchical in flavor space. Remarkably, an adequate hierarchy is naturally obtained if the RPV couplings are aligned with the SM flavor couplings~\cite{RPVMFV}. Indeed, under the assumption that there are no new flavor structures beyond the Yukawa couplings, lepton number violating couplings are forbidden while baryon number violating ones can be sizeable only when they involve the top flavor\footnote{When the neutrinos acquire a Majorana mass term, the lepton number violating couplings are permitted but are tuned by the tiny neutrino masses~\cite{RPVMFV}. Therefore, in the rest of this paper, it is understood that by R-parity violating couplings we always mean those violating baryon number only.}. In practice, to export in a controlled way the SM flavor hierarchies onto the MSSM flavor couplings, we use the minimal flavor violation (MFV) approach~\cite{MFVexp,MFVsym}, which is based on a well-defined symmetry principle~\cite{Georgi}.

In the present paper, our goal is to study the behavior of the minimally
flavor violating RPV couplings under the renormalization group (RG) evolution.
Indeed, if valid, the MFV hypothesis is likely to derive from a new flavor
dynamics taking place at a very high scale (see e.g.
Refs.~\cite{MFVdynamics1,MFVdynamics2,MFVdynamics3,MFVdynamics4,MFVdynamics5}%
), and it is crucial to check whether it survives down to the low scale. We
will see that this survival severely constrains the formulation of MFV, and in
particular the viable flavor symmetry group. At the same time, once these
constraints are in place, MFV is not only stable, it is even radiatively
reinforced through the evolution.

Before being able to delve into the numerics of the evolution, we must set up
the formalism. The first task is to construct a flavor-symmetric
reparametrization of the RPV couplings in terms of the Yukawa couplings. At
that stage, the RPV couplings need not satisfy MFV. Actually, this
reparametrization provides a unique way to specify fully generic RPV couplings
independently of the flavor basis chosen for the (s)quark fields, and thus
extends to the RPV sector the procedure proposed in Refs.~\cite{Mercolli:2009ns,NatSUSY}. This
is presented in Sections~\ref{FlavorBasis} and~\ref{Generic}.

With this tool in hand, the second step is to impose MFV. In the R-parity
conserving (RPC) sector, this is very easy: the reparametrization has to be
natural, hence must involve at most $\mathcal{O}(1)$
coefficients~\cite{Colangelo:2008qp}. In the RPV sector, however, we find that
this $\mathcal{O}(1)$ criterion is neither stable nor well-defined. The
reasons for this, and the conditions under which consistency is recovered, are
detailed in Section~\ref{MFVlimit}.

Once these initial steps are completed, the numerical study of the RG
evolution is undertaken in Section~\ref{RGEevol}. Special emphasis is laid on
the holomorphic implementation of MFV, as proposed in Ref.~\cite{CsakiGH11},
for which we prove several unique features, most notably the RG invariance. A
matrix identity based on Cayley-Hamilton identity is instrumental in this
proof, as well as in the construction of the reparametrization. It is derived
in Appendix~\ref{CHth}. Finally, the boundary conditions and mass spectrum of
the CMSSM-like scenario used to illustrate the behavior of the RG evolution
are collected in Appendix~\ref{CMSSM}.

\section{From generic to MFV couplings}

The MSSM flavor sector necessitates many parameters to be fully specified.
Even restricted to (s)quarks, there are 73 real constants, to which 18 complex
baryonic R-parity violating couplings must be added. Specifically, the
supersymmetric parameters occur in the superpotential,%
\begin{equation}
\mathcal{W}_{RPV}=-U^{I}\mathbf{Y}_{u}^{IJ}Q^{J}H_{u}+D^{I}\mathbf{Y}_{d}%
^{IJ}Q^{J}H_{d}+\frac{1}{2}\mathbf{Y}_{udd}^{IJK}U^{I}D^{J}D^{K}\;, \label{UDD1}%
\end{equation}
where $I,J,K=1,2,3$ denote flavor indices. Because of the understood
contraction over colors, $\mathbf{Y}_{udd}^{IJK}=-\mathbf{Y}_{udd}^{IKJ}$, and
R-parity violation is encoded into nine independent complex
couplings~\cite{Barbier04}. In addition, there are several soft-breaking terms
involving squark fields and with a priori non-trivial flavor structures,%
\begin{align}
\mathcal{L}_{soft}  &  =-\tilde{Q}^{\dagger I}(\mathbf{m}_{Q}^{2})^{IJ}%
\tilde{Q}^{J}-\tilde{U}^{I}(\mathbf{m}_{U}^{2})^{IJ}\tilde{U}^{\dagger
J}-\tilde{D}^{I}(\mathbf{m}_{D}^{2})^{IJ}\tilde{D}^{\dagger J}\nonumber\\
&  \;\;\;\;-\tilde{U}^{I}\mathbf{A}_{u}^{IJ}\tilde{Q}^{J}H_{u}+\tilde{D}%
^{I}\mathbf{A}_{d}^{IJ}\tilde{Q}^{J}H_{d}+\mathbf{A}_{udd}^{IJK}%
\tilde{U}^{I}\tilde{D}^{J}\tilde{D}^{K}+h.c.\;,
\end{align}
where $\mathbf{m}_{Q}^{2}$, $\mathbf{m}_{U}^{2}$, and $\mathbf{m}_{D}^{2}$ are
hermitian squark mass terms, while the R-parity violating couplings again
satisfy $\mathbf{A}_{udd}^{IJK}=-\mathbf{A}_{udd}^{IKJ}$ because of the color contraction.

At the same time, many of these parameters are constrained by the now precise
data from flavor physics. Assuming SUSY particles are not far heavier than
about one TeV, squark mixing cannot be large and R-parity violation must be
limited. In an attempt at systematically embedding these constraints, the
minimal flavor violation ansatz is particularly well
suited~\cite{MFVexp,MFVsym}. As a starting point towards MFV, let us first
construct an alternative parametrization of all these flavor couplings, using
as a guiding principle only the $U(3)^{3}$ flavor symmetry~\cite{Georgi} of
the MSSM (s)quark kinetic terms.

\subsection{Flavor-basis independence\label{FlavorBasis}}

When non-vanishing, all the flavor couplings break the $U(3)^{3}%
=U(3)_{Q}\otimes U(3)_{U}\otimes U(3)_{D}$ flavor symmetry, which means that
they vary when (s)quark undergo $U(3)$ rotations in flavor space. Since quark
mass terms originate from the Yukawa couplings $\mathbf{Y}_{u,d}$, this
freedom is in general used to bring all but either the up- or the down-type
left-handed quarks to their mass eigenstates. For example, when all quarks but
the $u_{L}$ are mass eigenstates, the (s)quark fields are rotated to the basis
where%
\begin{equation}
v_{u}\mathbf{Y}_{u}=\mathbf{m}_{u}\cdot\mathbf{V}_{CKM}\;,\;\;v_{d}%
\mathbf{Y}_{d}=\mathbf{m}_{d}\;,
\end{equation}
with $\mathbf{m}_{u,d}$ the diagonal quark mass matrices, $\mathbf{V}_{CKM}$
the CKM matrix, and $v_{u,d}$ the vacuum expectation values of $H_{u,d}^{0}$.
Equivalently, if $d_{L}$ is left out instead, then $v_{u}\mathbf{Y}%
_{u}=\mathbf{m}_{u}$ and $v_{d}\mathbf{Y}_{d}=\mathbf{m}_{d}\cdot
\mathbf{V}_{CKM}^{\dagger}$.

Obviously, performing the same unitary rotations on both quark and squark
fields redefines the $\mathbf{Y}_{udd}$ couplings as well as the soft-breaking
terms. For example, if the singular value decompositions for the Yukawa
couplings are denoted as $V_{R}^{u,d}\mathbf{Y}_{u,d}V_{L}^{u,d\dagger}$, then
the $V_{R}^{u,d}$ matrices find their way into $\mathbf{Y}_{udd}$ when
rotating $U\rightarrow UV_{R}^{u}$ and $D\rightarrow DV_{R}^{d}$. Except if a
flavor model is prescribed, $V_{R,L}^{u,d}$ are unknown matrices so there is a
numerical ambiguity in defining the whole flavor sector. More precisely,
non-universal soft-breaking terms are unambiguously defined only once the
Yukawa couplings in the same flavor basis are known. The situation is more
critical for the RPV couplings since they are never universal. In other words,
they always depend on the flavor basis.

This issue was discussed for the R-parity conserving MSSM in
Ref.~\cite{NatSUSY}. Let us briefly review the strategy proposed there to
circumvent the basis dependence. The trick is to write $\mathbf{m}_{Q,U,D}%
^{2}$ and $\mathbf{A}_{u,d}$ directly in terms of the Yukawa couplings. To
ensure their independence on the flavor basis, the $\mathbf{Y}_{u,d}$
couplings are treated as spurions, and the RPC soft-breaking terms are written
as manifestly $U(3)^{3}$ symmetric polynomial expansions in the spurions. Only
then are we certain that performing $U(3)^{3}$ rotations of the (s)quark
fields leaves invariant the expansion coefficients. Specifically, because both
$\mathbf{Y}_{u}$ and $\mathbf{Y}_{d}$ transform non-trivially under $U(3)_{Q}%
$, the most generic expansions are constructed by inserting in all possible
ways $SU(3)_{Q}$ octets, i.e., by inserting%
\begin{equation}
\mathbf{O}=\mathbf{1}\oplus\mathbf{X}_{u}\oplus\mathbf{X}_{d}\oplus
\mathbf{X}_{u}^{2}\oplus\mathbf{X}_{d}^{2}\oplus\{\mathbf{X}_{u}%
,\mathbf{X}_{d}\}\oplus i[\mathbf{X}_{u},\mathbf{X}_{d}]\oplus i[\mathbf{X}%
_{u}^{2},\mathbf{X}_{d}]\oplus i[\mathbf{X}_{u},\mathbf{X}_{d}^{2}%
]\;,\label{Octets}%
\end{equation}
where $\mathbf{X}_{u,d}\equiv\mathbf{Y}_{u,d}^{\dagger}\mathbf{Y}_{u,d}$ so
that $\mathbf{O}\rightarrow g_{Q}\mathbf{O}g_{Q}^{\dagger}$ under $g_{Q}\in
U(3)_{Q}$, and $\oplus$ denotes the presence of free expansion coefficients.
These nine independent terms are sufficient to project an arbitrary complex
matrix. They were obtained by first invoking the Cayley-Hamilton identity to
remove linearly dependent terms (see Appendix~\ref{CHth}), and then select the
nine terms involving the least number of Yukawa
insertions~\cite{Colangelo:2008qp}. In terms of these octets then,
\begin{subequations}
\label{RPCSoft}%
\begin{align}
\mathbf{m}_{Q}^{2}/m_{0}^{2} &  =a_{1}^{q}\,\mathbf{1}+a_{2}^{q}%
\,\mathbf{X}_{u}+a_{3}^{q}\,\mathbf{X}_{d}+a_{4}^{q}\,\mathbf{X}_{u}^{2}%
+a_{5}^{q}\,\mathbf{X}_{d}^{2}+a_{6}^{q}\,\{\mathbf{X}_{u},\mathbf{X}%
_{d}\}\nonumber\\
&  \;\;\;+b_{1}^{q}\,i[\mathbf{X}_{u},\mathbf{X}_{d}]+b_{2}^{q}\,i[\mathbf{X}%
_{u}^{2},\mathbf{X}_{d}]+b_{3}^{q}\,i[\mathbf{X}_{u},\mathbf{X}_{d}%
^{2}]\;\;,\;\frac{{}}{{}}\\
\mathbf{m}_{U,D}^{2}/m_{0}^{2} &  =a_{1}^{u,d}\,\mathbf{1}+\mathbf{Y}%
_{u,d}\,(a_{1}^{u,d}\,\mathbf{1}+a_{2}^{u,d}\,\mathbf{X}_{u}+a_{3}%
^{u,d}\,\mathbf{X}_{d}+a_{4}^{u,d}\,\mathbf{X}_{d,u}^{2}+a_{5}^{q}%
\,\{\mathbf{X}_{u},\mathbf{X}_{d}\})\,\mathbf{Y}_{u,d}^{\dagger}\nonumber\\
&  \;\;\;\;\;\;+\mathbf{Y}_{u,d}\,(b_{1}^{u,d}\,i[\mathbf{X}_{u}%
,\mathbf{X}_{d}]+b_{2}^{u,d}\,i[\mathbf{X}_{u}^{2},\mathbf{X}_{d}]+b_{3}%
^{u,d}\,i[\mathbf{X}_{u},\mathbf{X}_{d}^{2}])\,\mathbf{Y}_{u,d}^{\dagger
}\;\;,\;\frac{{}}{{}}\\
\mathbf{A}_{u,d}/A_{0} &  =\mathbf{Y}_{u,d}\,(c_{1}^{u,d}\,\mathbf{1}%
+c_{2}^{u,d}\,\mathbf{X}_{u}+c_{3}^{u,d}\,\mathbf{X}_{d}+c_{4}^{u,d}%
\,\mathbf{X}_{u}^{2}+c_{5}^{u,d}\,\mathbf{X}_{d}^{2}+c_{6}^{u,d}%
\,\{\mathbf{X}_{u},\mathbf{X}_{d}\}\nonumber\\
&  \;\;\;\;\;\;\;\;\;\;\;\;\;\;+c_{7}^{u,d}\,i[\mathbf{X}_{u},\mathbf{X}%
_{d}]+c_{8}^{u,d}\,i[\mathbf{X}_{u}^{2},\mathbf{X}_{d}]+c_{9}^{u,d}%
\mathbf{\,}i[\mathbf{X}_{u},\mathbf{X}_{d}^{2}])\;\;,
\end{align}
where both $A_{0}$ and $m_{0}$ set the soft-breaking scale. The crucial
observation is that the real coefficients $a_{i}^{q,u,d}$ and $b_{i}^{q,u,d}$
and the complex coefficients $c_{i}^{u,d}$ are independent of the flavor basis
in which the (s)quark fields are defined. They thus permit to unambiguously
parametrize the soft-breaking terms.

The goal in the next section is to construct the same kind of expansions for
the RPV couplings. But before detailing this construction, let us recall the
main properties or advantages of this procedure (for more details, see
Ref.~\cite{NatSUSY}):

\end{subequations}
\begin{enumerate}
\item As long as the coefficients are left free, any complex or hermitian
matrix can be projected onto the basis~(\ref{RPCSoft}). There are as many free
coefficients as there are free parameters~\cite{Nikolidakis08}.

\item The MFV hypothesis is immediate to formulate: it requires all the
coefficients to be of $\mathcal{O}(1)$ or smaller. By contrast, as
discussed in Ref.~\cite{Mercolli:2009ns}, the coefficients are in general much
larger than one whenever a new flavor structure not precisely aligned with the
Yukawa couplings is present.

\item These expansions can be defined at any scale, so the RG evolution can be
encoded into that of the expansion coefficients. In
Refs.~\cite{Paradisi:2008qh,Colangelo:2008qp}, the RPC coefficients were found
to exhibit infrared ``quasi''-fixed points. Interestingly, the RPV
coefficients also show such a behavior, as will be discussed later on.

\item CP-violating sources are naturally separated into the new phases
entering via the coefficients and those arising directly from the CKM phase
present in the Yukawa couplings.

\item In practice, when none of the coefficients is large, many terms in these
expansions can be dropped because $\mathbf{X}_{u,d}^{2}\approx\langle
\mathbf{X}_{u,d}\rangle\mathbf{X}_{u,d}$ with the flavor trace $\langle
\mathbf{X}_{u,d}\rangle\lesssim1$. In addition, when $\tan\beta$ is not large,
terms involving $\mathbf{X}_{d}$ are negligible compared to those involving
$\mathbf{X}_{u}$. In those cases, our procedure offers a simple
phenomenological parametrization for a fully realistic flavor sector.
\end{enumerate}

As advocated in Ref.~\cite{NatSUSY}, at least as long as no flavor model is
introduced, the procedure of fixing flavor couplings through their flavor
coefficients is far better than fixing them directly in some
arbitrary basis, and should be implemented in the available numerical codes.
In addition, experimental constraints can be translated as limits on the size
of the various coefficients. Only in this case can one draw definitive
basis-independent conclusions on the size of the new flavor couplings. In this
respect, all the current flavor constraints, including
EDM~\cite{Mercolli:2009ns}, flavor observables~\cite{NirIP10}, or the
extremely tight proton decay bounds~\cite{RPVMFV}, do allow for $\mathcal{O}%
(1)$ coefficients.

\subsection{Generic RPV couplings\label{Generic}}

Let us now construct the expansions for the RPV couplings. Given that a
generic $\mathbf{Y}_{udd}$ introduces nine arbitrary complex parameters, the
simplest polynomial expansions require nine independent terms. The strategy to
chose them is to first consider possible contractions with epsilon tensors.
This step was described in Ref.~\cite{RPVMFV}. Here, we consider only the
three simplest epsilon structures
\begin{subequations}
\begin{align}
(\mathbf{Y}_{udd}^{Q})^{IJK}  &  \sim\varepsilon^{LMN}\mathbf{Y}_{u}%
^{IL}\mathbf{Y}_{d}^{JM}\mathbf{Y}_{d}^{KN}+...\;,\label{BasicQ}\\
(\mathbf{Y}_{udd}^{D})^{IJK}  &  \sim\varepsilon^{LJK}(\mathbf{Y}%
_{u}\mathbf{Y}_{d}^{\dagger})^{IL}+...\;,\label{BasicD}\\
(\mathbf{Y}_{udd}^{U})^{IJK}  &  \sim\varepsilon^{IMN}(\mathbf{Y}%
_{d}\mathbf{Y}_{u}^{\dagger})^{JM}(\mathbf{Y}_{d}\mathbf{Y}_{u}^{\dagger
})^{KN}+...\;, \label{BasicU}%
\end{align}
where either the epsilon tensor of $SU(3)_{Q}$, $SU(3)_{D}$, or $SU(3)_{U}$ is
used. From this, the most general expansions are obtained by inserting in all
possible ways the $SU(3)_{Q}$ octet expansions of Eq.~(\ref{Octets}). At this
stage, because of the epsilon contractions, some redundant terms remain. The
final step is to remove them and identify the minimal set of nine independent
terms using the matrix identities derived in Appendix~\ref{CHth}, which
combine Cayley-Hamilton theorem with the definition of the determinant.

These identities permit to get rid of many terms. Take for example the
$\varepsilon^{LMN}\mathbf{Y}_{u}^{IL}\mathbf{Y}_{d}^{JM}\mathbf{Y}_{d}^{KN}$
structure. Any $SU(3)_{Q}$ octet insertion acting on $\mathbf{Y}_{d}$ can be
moved to act on $\mathbf{Y}_{u}$ using either Eq.~(\ref{Det1CHa}),%
\end{subequations}
\begin{equation}
\varepsilon^{LMN}\mathbf{Y}_{u}^{IL}[(\mathbf{Y}_{d}\mathbf{O})^{JM}%
\mathbf{Y}_{d}^{KN}+\mathbf{Y}_{d}^{JM}(\mathbf{Y}_{d}\mathbf{O}%
)^{KN}]=\varepsilon^{LMN}(\mathbf{Y}_{u}\left[  \langle\mathbf{O}%
\rangle-\mathbf{O}\right]  )^{IL}\mathbf{Y}_{d}^{JM}\mathbf{Y}_{d}^{KN}\;,
\label{Id1}%
\end{equation}
where the two terms on the left-hand side enforce $\mathbf{Y}_{udd}%
^{IJK}=-\mathbf{Y}_{udd}^{IKJ}$, or Eq.~(\ref{Det2CHb}),%
\begin{equation}
\varepsilon^{LMN}\mathbf{Y}_{u}^{IL}(\mathbf{Y}_{d}\mathbf{O})^{JM}%
(\mathbf{Y}_{d}\mathbf{O})^{KN}=\varepsilon^{LMN}(\mathbf{Y}_{u}%
[\mathbf{O}^{2}-\langle\mathbf{O}\rangle\mathbf{O}+\tfrac{1}{2}\langle
\mathbf{O}\rangle^{2}-\tfrac{1}{2}\langle\mathbf{O}^{2}\rangle])^{IL}%
\mathbf{Y}_{d}^{JM}\mathbf{Y}_{d}^{KN}\;, \label{Id2}%
\end{equation}
where $\mathbf{O}$ is an arbitrary complex matrix. The right-hand side retains
a manifestly $SU(3)_{Q}$ invariant form since $\mathbf{O}$ transforms as an
octet. Therefore, octets need to act on the $\mathbf{Y}_{u}$ factor only, and
the final set of nine terms can be chosen as (remember $\mathbf{X}_{u,d}%
\equiv\mathbf{Y}_{u,d}^{\dagger}\mathbf{Y}_{u,d}$)%
\begin{multline}
(\mathbf{Y}_{udd}^{Q})^{IJK}=\varepsilon^{LMN}(\mathbf{Y}_{u}(\lambda_{1}%
^{q}\mathbf{1}+\lambda_{2}^{q}\mathbf{X}_{u}+\lambda_{3}^{q}\mathbf{X}%
_{d}+\lambda_{4}^{q}\mathbf{X}_{u}^{2}+\lambda_{5}^{q}\mathbf{X}_{d}%
^{2}+\lambda_{6}^{q}\{\mathbf{X}_{u},\mathbf{X}_{d}\}\\
+\lambda_{7}^{q}i[\mathbf{X}_{u},\mathbf{X}_{d}]+\lambda_{8}^{q}%
i[\mathbf{X}_{u}^{2},\mathbf{X}_{d}]+\lambda_{9}^{q}i[\mathbf{X}%
_{u},\mathbf{X}_{d}^{2}]\mathbf{)})^{IL}\mathbf{Y}_{d}^{JM}\mathbf{Y}_{d}%
^{KN}\;,\;\;\;\;\;\; \label{LamdQ}%
\end{multline}
where $\lambda_{1,...,9}^{q}$ are nine free complex parameters. A similar
reduction can be done starting from Eq.~(\ref{BasicD}), leading to the
alternative basis%
\begin{multline}
(\mathbf{Y}_{udd}^{D})^{IJK}=\varepsilon^{LJK}(\mathbf{Y}_{u}(\lambda_{1}%
^{d}\mathbf{1}+\lambda_{2}^{d}\mathbf{X}_{u}+\lambda_{3}^{d}\mathbf{X}%
_{d}+\lambda_{4}^{d}\mathbf{X}_{u}^{2}+\lambda_{5}^{d}\mathbf{X}_{d}%
^{2}+\lambda_{6}^{d}\{\mathbf{X}_{u},\mathbf{X}_{d}\}\\
+\lambda_{7}^{d}i[\mathbf{X}_{u},\mathbf{X}_{d}]+\lambda_{8}^{d}%
i[\mathbf{X}_{u}^{2},\mathbf{X}_{d}]+\lambda_{9}^{d}i[\mathbf{X}%
_{u},\mathbf{X}_{d}^{2}]\mathbf{)Y}_{d}^{\dagger})^{IL}\;.\;\;\;\;\;\;
\label{LamdD}%
\end{multline}
Finally, for the last structure, Eq.~(\ref{BasicU}), all octet insertions but
those involving $\mathbf{Y}_{d}\mathbf{X}_{d}\mathbf{Y}_{u}^{\dagger}$ and
$\mathbf{Y}_{d}\mathbf{X}_{d}^{2}\mathbf{Y}_{u}^{\dagger}$ can be moved to the
first index, and we remain with 12 possible terms. This time, there seems to
be some latitude in the identification of the basis. For reasons that will be
detailed below, the best choice is to keep two such $\mathbf{X}_{d}$
insertions (which have to be antisymmetrized under $J\leftrightarrow K$):%
\begin{align}
(\mathbf{Y}_{udd}^{U})^{IJK}  &  =\varepsilon^{LMN}(\lambda_{1}^{u}%
\mathbf{1}+\mathbf{Y}_{u}(\lambda_{2}^{u}\mathbf{1}+\lambda_{4}^{u}%
\mathbf{X}_{u}+\lambda_{5}^{u}\mathbf{X}_{d}+\lambda_{7}^{u}\mathbf{X}_{d}%
^{2})\mathbf{Y}_{u}^{\dagger})^{IL}\;(\mathbf{Y}_{d}\mathbf{Y}_{u}^{\dagger
})^{JM}\;(\mathbf{Y}_{d}\mathbf{Y}_{u}^{\dagger})^{KN}\nonumber\\
&  \;\;\;\;\;+\varepsilon^{LMN}(\mathbf{Y}_{u}(\lambda_{8}^{u}\{\mathbf{X}%
_{u},\mathbf{X}_{d}\}+\lambda_{9}^{u}i[\mathbf{X}_{u},\mathbf{X}%
_{d}])\mathbf{Y}_{u}^{\dagger})^{IL}\;(\mathbf{Y}_{d}\mathbf{Y}_{u}^{\dagger
})^{JM}\;(\mathbf{Y}_{d}\mathbf{Y}_{u}^{\dagger})^{KN}\nonumber\\
&  \;\;\;\;\;+\varepsilon^{IMN}\lambda_{3}^{u}((\mathbf{Y}_{d}\mathbf{X}%
_{d}\mathbf{Y}_{u}^{\dagger})^{JM}(\mathbf{Y}_{d}\mathbf{Y}_{u}^{\dagger
})^{KN}+(\mathbf{Y}_{d}\mathbf{Y}_{u}^{\dagger})^{JM}(\mathbf{Y}_{d}%
\mathbf{X}_{d}\mathbf{Y}_{u}^{\dagger})^{KN})\nonumber\\
&  \;\;\;\;\;+\varepsilon^{IMN}\lambda_{6}^{u}(\mathbf{Y}_{d}\mathbf{X}%
_{d}\mathbf{Y}_{u}^{\dagger})^{JM}(\mathbf{Y}_{d}\mathbf{X}_{d}\mathbf{Y}%
_{u}^{\dagger})^{KN}\;,\; \label{LamdU}%
\end{align}
where the coefficients are ordered according to the number of Yukawa spurions.

The RPV soft-breaking term $\mathbf{A}_{udd}$ transforms exactly like
$\mathbf{Y}_{udd}$ under the $SU(3)^{3}$ symmetry, so admits the same
expansions, up to a prefactor $A_{0}$ setting the soft-breaking scale, and of
course a priori different coefficients. Therefore,
\begin{multline}
(\mathbf{A}_{udd}^{Q})^{IJK}=A_{0}\,\varepsilon^{LMN}(\mathbf{Y}_{u}%
(\kappa_{1}^{q}\mathbf{1}+\kappa_{2}^{q}\mathbf{X}_{u}+\kappa_{3}%
^{q}\mathbf{X}_{d}+\kappa_{4}^{q}\mathbf{X}_{u}^{2}+\kappa_{5}^{q}%
\mathbf{X}_{d}^{2}+\kappa_{6}^{q}\{\mathbf{X}_{u},\mathbf{X}_{d}\}\\
+\kappa_{7}^{q}i[\mathbf{X}_{u},\mathbf{X}_{d}]+\kappa_{8}^{q}i[\mathbf{X}%
_{u}^{2},\mathbf{X}_{d}]+\kappa_{9}^{q}i[\mathbf{X}_{u},\mathbf{X}_{d}%
^{2}]\mathbf{)})^{IL}\mathbf{Y}_{d}^{JM}\mathbf{Y}_{d}^{KN}\;,\;\;\;\;\;\;
\label{AuddQ}%
\end{multline}
and similarly for $\mathbf{A}_{udd}^{U,D}$.

\subsection{MFV limit for RPV couplings\label{MFVlimit}}

At this stage, one may wonder why three different bases, Eqs.~(\ref{LamdQ}),~(\ref{LamdD}),
and~(\ref{LamdU}), are constructed to parametrize $\mathbf{Y}_{udd}$. Indeed, any one of them is sufficient to project a completely arbitrary set of $\mathbf{Y}_{udd}^{IJK}$ couplings.
Generalizing, it is clear that there is an infinity of equally valid bases of nine terms, at least from a mathematical point of view.

Though this is indeed true when these bases are just meant to parametrize
generic couplings, the situation changes when MFV is enforced. Indeed, we must
make sure that the MFV limit is stable and well-defined; the second property
quoted in Section~\ref{FlavorBasis}. More precisely, if a flavor coupling is
expressed as a combination of Yukawa spurions with the adequate symmetry
properties and $\mathcal{O}(1)$ coefficients, then by definition it satisfies
the MFV requirement. Thus, once projected on a specific choice of basis, it
must give back $\mathcal{O}(1)$ coefficients only. This is trivial if that
particular combination of Yukawa spurions is part of the basis, but not
automatic otherwise, as we now explore.

\subsubsection{Internal stability of the epsilon contractions}

Within a given basis, i.e., for a given epsilon structure, the stability is
ensured by the systematic use of the Cayley-Hamilton theorem. For example, if
MFV holds, then $\langle\mathbf{O}\rangle$ in Eq.~(\ref{Id1}) and~(\ref{Id2})
is at most of $\mathcal{O}(1)$, hence can be absorbed into the coefficients
without upsetting their scaling~\cite{Colangelo:2008qp}. Therefore, both the
$\mathbf{Y}_{udd}^{Q}$ and $\mathbf{Y}_{udd}^{D}$ bases are internally
consistent. On the other hand, the $\mathbf{Y}_{udd}^{U}$ basis must contain
the $\lambda_{3}^{u}$ and $\lambda_{6}^{u}$ terms instead of for example
$\varepsilon^{LMN}(\mathbf{Y}_{u}[\mathbf{X}_{u,d}^{2},\mathbf{X}%
_{d,u}]\mathbf{Y}_{u}^{\dagger})^{IL}(\mathbf{Y}_{d}\mathbf{Y}_{u}^{\dagger
})^{JM}(\mathbf{Y}_{d}\mathbf{Y}_{u}^{\dagger})^{KN}$. If $\lambda_{3}^{u}$
and $\lambda_{6}^{u}$ were not part of the $\mathbf{Y}_{udd}^{U}$ basis, the
other terms could not reproduce them with only $\mathcal{O}(1)$ coefficients
because there is no matrix identity relating them. The converse holds though:
the $\varepsilon^{LMN}(\mathbf{Y}_{u}[\mathbf{X}_{u,d}^{2},\mathbf{X}%
_{d,u}]\mathbf{Y}_{u}^{\dagger})^{IL}(\mathbf{Y}_{d}\mathbf{Y}_{u}^{\dagger
})^{JM}(\mathbf{Y}_{d}\mathbf{Y}_{u}^{\dagger})^{KN}$ structures are so
suppressed numerically that no large coefficients are generated when projected
on the $\mathbf{Y}_{udd}^{U}$ basis of Eq.~(\ref{LamdU}).

\subsubsection{Incompatibility between epsilon contractions}

The stability of MFV within a given basis can be ensured, but not that between
the bases with different epsilon contractions. Consider for example the
identity:%
\begin{equation}
\varepsilon^{LMN}(\mathbf{Y}_{u}\mathbf{Y}_{d}^{\dagger}\mathbf{Y}_{d}%
)^{IL}\mathbf{Y}_{d}^{JM}\mathbf{Y}_{d}^{KN}=\det(\mathbf{Y}_{d}%
)\varepsilon^{LJK}(\mathbf{Y}_{u}\mathbf{Y}_{d}^{\dagger})^{IL}\;.
\end{equation}
It shows that projecting the $\lambda_{1}^{d}$ structure of the $\mathbf{Y}%
_{udd}^{D}$ basis on the $\mathbf{Y}_{udd}^{Q}$ basis just produces the
$\lambda_{3}^{q}$ term, but that $\lambda_{3}^{q}=\lambda_{1}^{d}%
/\det(\mathbf{Y}_{d})$. With $\det(\mathbf{Y}_{d})\approx10^{-10}\tan^{3}%
\beta$, it is clear that both $\lambda_{1}^{d}$ and $\lambda_{3}^{q}$ cannot
be simultaneously of $\mathcal{O}(1)$. Thus, what is MFV for one basis is not
necessarily MFV for another basis.

At this stage, there are two possible ways to restore a well-defined MFV principle. Either we combine terms from the three bases to construct a fully general one, or we constrain the possible $U(1)$ breakings. For example, if only $U(1)_{Q}$ is broken, then the $\mathbf{Y}_{udd}^{Q}$ basis suffices. Indeed, once $U(1)_{D}$ and $U(1)_{U}$ are enforced, all the terms of the $\mathbf{Y}_{udd}^{D}$ and $\mathbf{Y}_{udd}^{U}$ bases are forbidden, since they involve an epsilon tensor acting in either $SU(3)_{D}$ or $SU(3)_{U}$. This latter alternative will be followed here, because allowing for the simultaneous presence of different $U(1)$-breaking terms would cause also other difficulties, as detailed below.

\subsubsection{Compatibility with the R-parity conserving MFV expansions}

When constructing the expansions of the soft-breaking terms, Eq.~(\ref{RPCSoft}), the invariance under $U(3)^{3}$ is enforced. In principle, if the invariance under $SU(3)^{3}$ is imposed instead, additional terms should occur in their expansions, like for example%
\begin{equation}
(\mathbf{m}_{D}^{2})^{IJ}/m_{0}^{2}\ni\varepsilon^{LMN}\mathbf{Y}_{u}%
^{AL}\mathbf{Y}_{d}^{IM}\mathbf{Y}_{d}^{KN}\times\varepsilon^{RJK}%
(\mathbf{Y}_{d}\mathbf{Y}_{u}^{\dagger})^{RA}\;, \label{Corr1}%
\end{equation}
where $\varepsilon^{LMN}$ breaks $U(1)_{Q}$ and $\varepsilon^{RJK}$ breaks
$U(1)_{D}$, or%
\begin{equation}
(\mathbf{m}_{D}^{2})^{IJ}/m_{0}^{2}\ni\varepsilon^{LMN}\mathbf{Y}_{u}%
^{AL}\mathbf{Y}_{d}^{IM}\mathbf{Y}_{d}^{KN}\times\varepsilon^{RST}%
\mathbf{Y}_{u}^{\dagger RA}\mathbf{Y}_{d}^{\dagger SJ}\mathbf{Y}_{d}^{\dagger
TK}=\varepsilon^{LMN}\varepsilon^{RST}\mathbf{Y}_{d}^{IL}\mathbf{X}_{u}%
^{RM}\mathbf{X}_{d}^{SN}\mathbf{Y}_{d}^{\dagger TJ}\;, \label{Corr2}%
\end{equation}
where both epsilons break $U(1)_{Q}$. In this latter case, having two epsilons
acting in the same $SU(3)$ actually preserves the corresponding $U(1)$
symmetry, so this term must be redundant with those already present in
Eq.~(\ref{RPCSoft}). This can be checked explicitly by simplifying the
epsilon contractions while maintaining the flavor symmetry manifest as%
\begin{equation}
\varepsilon^{LMN}\varepsilon^{RST}\mathbf{Y}_{d}^{IL}\mathbf{X}_{u}%
^{RM}\mathbf{X}_{d}^{SN}\mathbf{Y}_{d}^{\dagger TJ}=(\mathbf{Y}_{d}%
(\{\mathbf{X}_{u},\mathbf{X}_{d}\}-\langle\mathbf{X}_{d}\rangle\mathbf{X}%
_{u}-\langle\mathbf{X}_{u}\rangle\mathbf{X}_{d}+\langle\mathbf{X}_{d}%
\rangle\langle\mathbf{X}_{u}\rangle-\langle\mathbf{X}_{u}\mathbf{X}_{d}%
\rangle)\mathbf{Y}_{d}^{\dagger})^{IJ}\;.
\end{equation}
On the contrary, the term of Eq.~(\ref{Corr1}) does not match those already
present in Eq.~(\ref{RPCSoft}). Even worse, if projected onto the MFV basis of
Eq.~(\ref{RPCSoft}), it generates large non-MFV coefficients. So, if one
insists on the pure $SU(3)^{3}$ invariance, with all the $U(1)$ simultaneously
broken, the usual MFV basis for the R-parity conserving soft-breaking terms
has to be extended.

It should be stressed that this is not just a matter of principle. Through the
RG evolution, the soft-breaking terms receive corrections from the RPV
couplings. For example, the one-loop $\beta$ function of $\mathbf{m}_{D}^{2}$
contains~\cite{RPVRGE1, RPVRGE2}%
\begin{equation}
(\beta_{\mathbf{m}_{D}^{2}})^{IJ}\ni\mathbf{Y}_{udd}^{\dagger AIB}%
\mathbf{Y}_{udd}^{ACB}(\mathbf{m}_{D}^{2})^{CJ}\;.
\end{equation}
Therefore, if $\mathbf{Y}_{udd}$ or $\mathbf{A}_{udd}$ contain epsilon tensors
acting in different $SU(3)$ spaces, terms similar to that in Eq.~(\ref{Corr1})
will occur. In that case, MFV would only be maintained through the RG
evolution provided additional terms are included in the expansions of
Eq.~(\ref{RPCSoft}). For the time being, we prefer not to follow that route.
We thus stick to the terms in Eq.~(\ref{RPCSoft}), but must allow for only a
single flavored $U(1)$ to be broken when constructing the expansions for the
RPV couplings $\mathbf{Y}_{udd}$ and $\mathbf{A}_{udd}$.

\subsubsection{U(1) phases and Yukawa background values}

At the beginning of the previous section, we stated that it is always possible
to perform $U(3)^{3}$ rotations to reach a basis where, e.g., $v_{u}%
\mathbf{Y}_{u}=\mathbf{m}_{u}\cdot\mathbf{V}_{CKM}$ and $v_{d}\mathbf{Y}%
_{d}=\mathbf{m}_{d}$. However, only the invariance under $SU(3)^{3}$ was used
in the construction of the $\mathbf{Y}_{udd}$ expansion. Because of this
mismatch, these expansions may not fully fulfill their role of rendering
$\mathbf{Y}_{udd}$ independent of the flavor basis: the unknown phases
corresponding to the broken $U(1)$ affect the coefficients.

Let us be more precise. The singular value decompositions of the Yukawa
couplings are
\begin{equation}
V_{R}^{u,d}\mathbf{Y}_{u,d}V_{L}^{u,d\dagger}=\mathbf{m}_{u,d}/v_{u,d}%
\;,\;\;\mathbf{V}_{CKM}=V_{L}^{u}V_{L}^{d\dagger}\;,
\end{equation}
where $\mathbf{m}_{u,d}$ are diagonal and positive-definite, $V_{R}^{u,d}\in U(3)_{U,D}$ and $V_{L}^{u,d}\in U(3)_{Q}$. Clearly, $V_{L}^{u}$ and $V_{R}^{u}$ are defined up to a diagonal matrix of phases $D_{u}=\operatorname{diag}(\exp(i\alpha_{1}^{u}),\exp(i\alpha_{2}^{u}),\exp(i\alpha_{3}^{u}))$, since $\mathbf{m}_{u}=D^{\dagger}\mathbf{m}_{u}D$, and similarly for the down sector. Requiring that $\mathbf{m}_{u,d}$ have real and positive entries only and that $\mathbf{V}_{CKM}$ is conventionally phased remove five linear combinations of the six phases. At that stage, $\det(V_{L}^{u})=\det(V_{L}^{d})\neq\det(V_{R}^{u})\neq\det(V_{R}^{d})$ in general, and they all depend on the remaining sixth phase. So, it is always possible to force either $V_{L}^{u,d}\in SU(3)_{Q}$, or $V_{R}^{u}\in SU(3)_{U}$, or $V_{R}^{d}\in SU(3)_{D}$, but not more than that. Said differently, we need at least an exact $SU(3)_{Q}\otimes U(3)_{U}\otimes U(3)_{D}$, $U(3)_{Q}\otimes SU(3)_{U}\otimes U(3)_{D}$, or $U(3)_{Q}\otimes U(3)_{U}\otimes SU(3)_{D}$ flavor symmetry\footnote{The fact that these $U(1)$ transformations are anomalous is inessential here, and the rephasing used to remove one of the $U(1)$ rotations does not affect the induced correction to the $\theta$ term since it depends only on $\mathbf{Y}_{u,d}$.} to reach the background values $v_{u}\mathbf{Y}_{u}=\mathbf{m}_{u}\cdot\mathbf{V}_{CKM}$ and $v_{d}\mathbf{Y}_{d}=\mathbf{m}_{d}$ (or equivalently, $v_{u}\mathbf{Y}_{u}=\mathbf{m}_{u}$ and $v_{d}\mathbf{Y}_{d}=\mathbf{m}_{d}\cdot\mathbf{V}_{CKM}^{\dagger}$).

Since two out of the three $U(1)$s of $U(3)^{3}$ have to remain exact, only
the epsilon tensor of a single $SU(3)$ can occur in the expansions of the RPV
couplings. This constraint prevents the phases of the expansion coefficients
from depending on the flavor basis. For example, if both $\mathbf{Y}_{udd}%
^{Q}$ and $\mathbf{Y}_{udd}^{D}$ are present, then the phases of the
$\mathbf{Y}_{udd}^{Q}$ coefficients depend on $\arg(\det(V_{L}^{u,d}))$ and
those of $\mathbf{Y}_{udd}^{D}$ on $\arg(\det(V_{R}^{d}))$, but both
$\arg(\det(V_{L}^{u,d}))$ and $\arg(\det(V_{R}^{d}))$ cannot be set to zero in
general. Therefore, for this and the other reasons discussed above, we will restrict
our attention to scenarios where only a single $U(1)$ is broken in the rest of
the paper.

\section{Renormalization group evolution\label{RGEevol}}

In the previous section, we have seen that simply asking for MFV to have a
chance to remain valid through the running brings a strong restriction on its
formulation. Only one $U(1)$ can be broken at a time. Consequently, there are
only three possible patterns of hierarchies for the RPV couplings when MFV is
valid, and those depend only on $\tan\beta$. For example, with $\tan\beta=10$,
both $\mathbf{Y}_{udd}$ and $\mathbf{A}_{udd}/A_{0}$ scale as in
Table~\ref{Hierarchies}.

\begin{table}[t]
\centering                     $%
\begin{tabular}
[c]{ccc}\hline
Broken $U(1)_{Q}$ & Broken $U(1)_{D}$ & Broken $U(1)_{U}$\\\hline
$\;\;\;\;\;%
\begin{array}
[c]{ccc}%
ds\;\; & \;\;sb\;\; & \;\;db
\end{array}
$ & $%
\begin{array}
[c]{ccc}%
ds\;\; & \;\;sb\;\; & \;\;db
\end{array}
$ & $%
\begin{array}
[c]{ccc}%
ds\;\; & \;\;sb\;\; & \;\;db
\end{array}
$\\
$%
\begin{array}
[c]{c}%
u\\
c\\
t
\end{array}
\left(
\begin{array}
[c]{ccc}%
10^{-14} & 10^{-9} & 10^{-11}\\
10^{-12} & 10^{-7} & 10^{-7}\\
10^{-7} & 10^{-6} & 10^{-6}%
\end{array}
\right)  $ & $\left(
\begin{array}
[c]{ccc}%
10^{-9} & 10^{-9} & 10^{-9}\\
10^{-5} & 10^{-7} & 10^{-5}\\
0.1 & 10^{-6} & 10^{-4}%
\end{array}
\right)  $ & $\left(
\begin{array}
[c]{ccc}%
10^{-12} & 10^{-6} & 10^{-8}\\
10^{-13} & 10^{-9} & 10^{-10}\\
10^{-14} & 10^{-13} & 10^{-14}%
\end{array}
\right)  $\\\hline
\end{tabular}
\ \ \ $ \caption{Typical hierarchies for the modulus of the $\mathbf{Y}_{udd}$ couplings (in
the superCKM basis) at the low scale under MFV with either $U(1)_{Q}$, $U(1)_{D}$, or $U(1)_{U}$ broken, and when $\tan\beta=10$. Because $\mathbf{Y}_{udd}^{IJK}$ is antisymmetric under $J\leftrightarrow K$, its entries can be put in a $3\times3$ matrix form with $I=u,c,t$ and
$JK=ds,sb,db$. Hierarchies for the RPV trilinear coupling $\mathbf{A}_{udd}/A_{0}$ are similar.}%
\label{Hierarchies}%
\end{table}

In the present section, we investigate in details the evolution of the coefficients. We start with the broken $U(1)_{Q}$ scenario, whose main interest is to cover the special case of holomorphic MFV~\cite{CsakiGH11}. As a result, we will see that this scenario has several unique properties, not
shared by any other couplings under MFV. By contrast, the behavior of the broken $U(1)_{D}$ or
$U(1)_{U}$ scenarios is more in line with that of the RPC soft-breaking terms~\cite{Paradisi:2008qh,Colangelo:2008qp}. This will be illustrated for the broken $U(1)_{D}$ case only. A detailed analysis of the $U(1)_{U}$ case is not very useful since it is similar. In addition, looking at Table~\ref{Hierarchies}, this scenario is much less interesting phenomenologically. First, the (s)top couplings are the largest when $U(1)_{D}$ is broken, but never exceed $\mathcal{O}(10^{-13})$ for $U(1)_{U}$. The same-sign top quark signals~\cite{Allanach:2012vj,AsanoRS13,DurieuxEtAl,BergerPST13,DurieuxS13,Batell:2013zwa,Saelim:2013gea,Allanach:2013qna,Bhattacherjee:2013tha} at the LHC would thus essentially disappear, and be replaced by the more challenging two or three light-jet resonances. Second, the couplings involving the up quark are the largest when $U(1)_{U}$ is broken, hence the sparticles have to be heavier to pass the current bounds on the proton lifetime or neutron oscillation. Finally, note that in all three scenarios some RPV couplings are tiny. This can indirectly constrain the supersymmetric mass spectrum because the squark lifetimes have to be short enough to circumvent R-hadron signatures~\cite{CsakiGH11,DurieuxS13}.

The RG evolution of the $\mathbf{A}_{udd}$ couplings will also be discussed
for the broken $U(1)_{Q}$ and broken $U(1)_{D}$ scenarios, though briefly.
Indeed, the impact of $\mathbf{A}_{udd}$ is very limited phenomenologically.
Whenever an $\mathbf{A}_{udd}$ coupling is large, the corresponding
$\mathbf{Y}_{udd}$ coupling is also large. So, if a squark can decay into two
other squarks through $\mathbf{A}_{udd}$, it can also decay to the
corresponding quarks through $\mathbf{Y}_{udd}$ with a larger available
phase-space. For this reason, except maybe for a slight reduction in the RPV
branching ratios to quark final states, even a relatively large $\mathbf{A}_{udd}$ coupling does not significantly affect the RPV signatures at the LHC.

Throughout this section, to illustrate the evolution of the RPV expansion
coefficients in a realistic setting, we use the CMSSM-like parameter point
described in Appendix~\ref{CMSSM}. We select the boundary conditions at the
GUT scale so that, in the RPC case, the Higgs boson mass is close to 125 GeV.
The impact of the RPV couplings on the particle spectrum is in general limited
since most RPV couplings are very suppressed, hence will be neglected here.

\subsection{RG invariance of MFV holomorphy}

The holomorphic restriction of MFV proposed in Ref.~\cite{CsakiGH11} originates from the hypothesis that the flavor symmetry is dynamical at some scale $M_{\mathrm{Flavor}}$. There, the Yukawa spurions would either be true dynamical fields, or they would be directly related to those of this unknown flavor dynamics. At the same time, supersymmetry requires the superpotential to be holomorphic, so $\mathbf{Y}_{udd}$ must be insensitive to $\mathbf{Y}_{u}^{\dagger}$ and $\mathbf{Y}_{d}^{\dagger}$ above the scale $M_{\mathrm{Flavor}}$. The most general flavor-symmetric expansion is then very simple, since there is only one way to write $\mathbf{Y}_{udd}$ in terms of $\mathbf{Y}_{u}$ and $\mathbf{Y}_{d}$:%
\begin{equation}
\mathbf{Y}_{udd}^{IJK}=\lambda\,\varepsilon^{LMN}\mathbf{Y}_{u}^{IL}%
\mathbf{Y}_{d}^{JM}\mathbf{Y}_{d}^{KN}\;.\label{Holo}%
\end{equation}
The holomorphic restriction thus respects MFV under the $SU(3)_{Q}\otimes U(3)_{U}\otimes U(3)_{D}$ flavor group. With only $U(1)_{Q}$ broken, it respects all the requirements discussed in the previous section and MFV is
stable and well defined.

However, the scale $M_{\mathrm{Flavor}}$ at which holomorphy is imposed could
be very high. Even if MFV is in itself stable, whether holomorphy is a
reasonable approximation at the low scale is not obvious. Indeed, the RG
equations of the Yukawa and $\mathbf{Y}_{udd}$ couplings are coupled (we
follow the notations of Ref.~\cite{RPVRGE1, RPVRGE2}, but for a slight change of
conventions in the indices):%
\begin{align}
\frac{d}{dt}\mathbf{Y}_{u}^{IJ} &  =\mathbf{Y}_{u}^{KJ}\gamma_{U^{K}}^{U^{I}%
}+\mathbf{Y}_{u}^{IJ}\gamma_{H_{2}}^{H_{2}}+\mathbf{Y}_{u}^{IK}\gamma_{Q^{K}%
}^{Q^{J}}\;,\\
\frac{d}{dt}\mathbf{Y}_{d}^{IJ} &  =\mathbf{Y}_{d}^{KJ}\gamma_{D^{K}}^{D^{I}%
}+\mathbf{Y}_{d}^{IJ}\gamma_{H_{1}}^{H_{1}}+\mathbf{Y}_{d}^{IK}\gamma_{Q^{K}%
}^{Q^{J}}\;,\\
\frac{d}{dt}\mathbf{Y}_{udd}^{IJK} &  =\mathbf{Y}_{udd}^{IJL}\gamma_{D^{L}%
}^{D^{K}}+\mathbf{Y}_{udd}^{ILK}\gamma_{D^{L}}^{D^{J}}+\mathbf{Y}_{udd}%
^{LJK}\gamma_{U^{L}}^{U^{I}}\;,
\end{align}
where $t=\log Q^{2}$. At one loop, $\gamma_{U^{J}}^{U^{I}}$, $\gamma_{D^{J}}^{D^{I}}$, and $\gamma_{Q^{J}}^{Q^{I}}$ all involve ``non-holomorphic'' spurion insertions. For example, $\gamma_{Q^{J}}^{Q^{I}}$ contains $\mathbf{Y}_{u}^{\dagger}\mathbf{Y}_{u}$ and $\mathbf{Y}_{d}^{\dagger}\mathbf{Y}_{d}$ terms. The consequence for the soft-breaking terms is well-known: even starting from universal squark masses $\mathbf{m}_{Q}^{2}=\mathbf{m}_{U}^{2}=\mathbf{m}_{D}^{2}=m_{0}^{2}\mathbf{1}$ at the unification scale, the whole series of coefficients in Eq.~(\ref{RPCSoft}) end up non-zero at the low scale. One would expect the same to happen for the $\mathbf{Y}_{udd}$ coupling: the whole series of coefficients in Eq.~(\ref{LamdQ}) would appear at the low scale.

Interestingly, the holomorphy of $\mathbf{Y}_{udd}$ holds at all scale because
all these non-holomorphic effects precisely cancel out. This can be checked
analytically:%
\begin{align}
\frac{d}{dt}\mathbf{Y}_{udd}^{IJK} &  =\frac{d}{dt}(\lambda\varepsilon
^{LMN}\mathbf{Y}_{u}^{IL}\mathbf{Y}_{d}^{JM}\mathbf{Y}_{d}^{KN})\nonumber\\
&  =\lambda\varepsilon^{LMN}\left(  \frac{d\ln\lambda}{dt}\mathbf{Y}_{u}%
^{IL}\mathbf{Y}_{d}^{JM}\mathbf{Y}_{d}^{KN}+\frac{d\mathbf{Y}_{u}^{IL}}%
{dt}\mathbf{Y}_{d}^{JM}\mathbf{Y}_{d}^{KN}+\mathbf{Y}_{u}^{IL}%
\frac{d\mathbf{Y}_{d}^{JM}}{dt}\mathbf{Y}_{d}^{KN}+\mathbf{Y}_{u}%
^{IL}\mathbf{Y}_{d}^{JM}\frac{d\mathbf{Y}_{d}^{KN}}{dt}\right)  \nonumber\\
&  =\mathbf{Y}_{udd}^{IJK}\left(  \frac{d\ln\lambda}{dt}+\gamma_{Q^{P}}%
^{Q^{P}}+\gamma_{H_{2}}^{H_{2}}+2\gamma_{H_{1}}^{H_{1}}\right)  +\frac{d}%
{dt}\mathbf{Y}_{udd}^{IJK}\;.
\end{align}
To reach the last line, we have used the matrix identity of Eq.~(\ref{Det1CHa}) in the form%
\begin{equation}
\gamma_{Q^{P}}^{Q^{P}}\varepsilon^{LMN}=\varepsilon^{PMN}\gamma_{Q^{L}}%
^{Q^{P}}+\varepsilon^{LPN}\gamma_{Q^{M}}^{Q^{P}}+\varepsilon^{LMP}%
\gamma_{Q^{N}}^{Q^{P}}\;.
\end{equation}
Therefore, the whole evolution of the holomorphic $\mathbf{Y}_{udd}$ can be encoded into a single coefficient:%
\begin{equation}
\frac{d\lambda}{dt}=-\lambda\beta_{\lambda}\;,\;\beta_{\lambda}=\gamma_{Q^{P}%
}^{Q^{P}}+\gamma_{H_{2}}^{H_{2}}+2\gamma_{H_{1}}^{H_{1}}\;.\label{RGEl}%
\end{equation}
The linear dependence of $d\lambda/dt$ over $\lambda$ ensures the RG
invariance of $\lambda=0$, when R-parity is unbroken. The beta function
$\beta_{\lambda}$ involves only purely left-handed anomalous terms: its sole
role is to compensate for the left-handed evolutions of the Yukawa couplings,
since $\mathbf{Y}_{udd}$ evolves according to right-handed anomalous terms
only. This explains the mechanism behind the RG invariance\footnote{It is
important to realize that while MFV holomorphy is an RG invariant property
for $\mathbf{Y}_{udd}$, these couplings are far from invariant numerically.
Not only is the coefficient evolving, but the Yukawa couplings on which
$\mathbf{Y}_{udd}$ is defined are themselves scale-dependent.} of the
$\varepsilon^{LMN}\mathbf{Y}_{u}^{IL}\mathbf{Y}_{d}^{JM}\mathbf{Y}_{d}^{KN}$
structure: only that term both brings in just the required combination of
right-handed quark anomalous dimensions, and at the same time leaves the rest
as a pure flavor trace. No other structure could be RG invariant.

At the one-loop order, the beta function of the coefficient $\lambda$
is~\cite{RPVRGE1, RPVRGE2}%
\begin{equation}
\beta_{\lambda}=\frac{1}{32\pi^{2}}(4\langle\mathbf{Y}_{u}^{\dagger}%
\mathbf{Y}_{u}\rangle+7\langle\mathbf{Y}_{d}^{\dagger}\mathbf{Y}_{d}%
\rangle+2\langle\mathbf{Y}_{e}^{\dagger}\mathbf{Y}_{e}\rangle-g_{1}^{2}%
-9g_{2}^{2}-8g_{3}^{2})\;,
\end{equation}
where $g_{1}$, $g_{2}$, and $g_{3}$ are the $U(1)_{Y}$, $SU(2)_{L}$, and
$SU(3)_{C}$ gauge couplings (with the $SU(5)$ normalization for the
hypercharge). The leading order RG equation of $\lambda$ can easily be solved.
Indeed, the evolution of the Yukawa couplings depends quadratically on
$\mathbf{Y}_{udd}$, whose maximal entry in the holomorphic case is about $\lambda\times 10^{-4}$ when $\tan \beta \approx 50$. Except for very large non-MFV values of the
coefficient, the impact of $\mathbf{Y}_{udd}$ on $\mathbf{Y}_{u,d}$ is
completely negligible. So, the ratio between the coefficients at the GUT and
SUSY scale is immediately found once the RPC evolution of the Yukawa and gauge
couplings is known,%
\begin{equation}
\frac{\lambda\lbrack M_{\mathrm{SUSY}}]}{\lambda\lbrack M_{\mathrm{GUT}}%
]}=\exp\left\{  -\int_{\log M_{\mathrm{SUSY}}^{2}}^{\log M_{\mathrm{GUT}}^{2}%
}\beta_{\lambda}(t)dt\right\}  \approx\exp\left\{  -\frac{1}{32\pi^2}\int_{\log
M_{\mathrm{SUSY}}^{2}}^{\log M_{\mathrm{GUT}}^{2}}(4y_{t}^{2}-9g_{2}%
^{2}-8g_{3}^{2})dt\right\}_{\mathrm{RPC}}  \;.\label{SolHolo}%
\end{equation}
Numerically, the right-hand side has only a very weak dependence on the rest
of the MSSM parameters, essentially through threshold corrections. Though the
sensitivity is a bit enhanced by the exponential, we find that with
$M_{\mathrm{SUSY}}\approx1$~TeV, the ratio is quite stable, varying within
$1/5$ and $1/4$. For the scenario of Appendix~\ref{CMSSM}, we get
$\lambda\lbrack M_{\mathrm{SUSY}}]/\lambda\lbrack M_{\mathrm{GUT}}]=0.2205$. Note though that the $\mathbf{Y}_{udd}$ couplings are nevertheless larger at the low scale because the decrease of $\lambda$ is more than compensated by the increase of the Yukawa couplings.

The other R-parity conserving parameters are also insensitive to
$\mathbf{Y}_{udd}$ when $\lambda$ is of $\mathcal{O}(1)$, as well as to the
RPV soft-breaking term $\mathbf{A}_{udd}$ when its overall mass scale is of
the same order as that of $\mathbf{A}_{u}$ and $\mathbf{A}_{d}$ (i.e., all are
tuned by the same $A_{0}$ parameter, see Eq.~(\ref{RPCSoft})). Indeed, under
these assumptions and since holomorphy does not apply to soft breaking terms,
$\mathbf{A}_{udd}$ admits an expansion of the form in Eq.~(\ref{AuddQ}) with
the nine $\kappa_{i}^{q}$ coefficients of $\mathcal{O}(1)$ or less.
Numerically, all the $\mathbf{A}_{udd}$ couplings are then very suppressed
compared to $\mathbf{A}_{u}$ and $\mathbf{A}_{d}$.

\begin{figure}[t]
\centering                  \includegraphics[width=15.5cm]{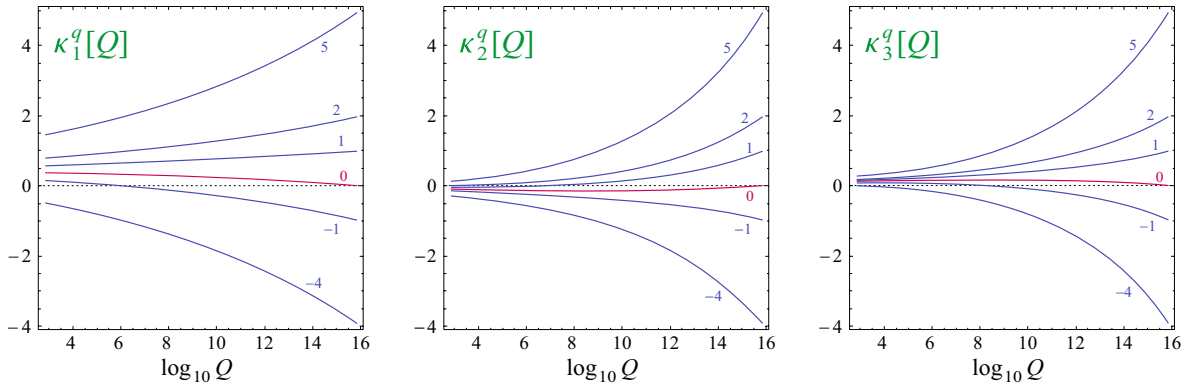}
\caption{Evolution of the leading coefficients for the trilinear couplings $\mathbf{A}_{udd}$ in the holomorphic scenario, with $\lambda\lbrack M_{\mathrm{GUT}}]=1$. The GUT scale boundary conditions are set as described in Appendix~\ref{CMSSM}, together with $\kappa_{i}^{q}[M_{\mathrm{GUT}}]=\kappa\delta_{ia}$ with $\kappa=\{-4,-1,0,1,2,5\}$ and $a=1\ $(left), $a=2$ (middle), and $a=3$ (right). The central curves (in red) correspond to the purely radiative generation, with the values at $M_{\mathrm{SUSY}}=1$~TeV given in Eq.~(\ref{kappaQval}).}%
\label{kappaQ}%
\end{figure}

On the other hand, $\mathbf{Y}_{udd}$ does affect $\mathbf{A}_{udd}$
significantly. Even starting from $\mathbf{A}_{udd}^{IJK}[M_{\mathrm{GUT}}%
]=0$, it is radiatively induced entirely through $\mathbf{Y}_{udd}$ and end up
sizeable at the low scale. For instance, using the benchmark point of
Appendix~\ref{CMSSM} with $\lambda\lbrack M_{\mathrm{GUT}}]=1$ and $\kappa
_{i}^{q}[M_{\mathrm{GUT}}]=0$, we find the MFV coefficients for $\mathbf{A}%
_{udd}[M_{\mathrm{SUSY}}\approx1~$TeV$]$ to be%
\begin{equation}
\kappa_{1,2,3}^{q}[M_{\mathrm{SUSY}}]=(0.36,\;-0.12,\;0.12)\;,\;\;\kappa
_{4,...,9}^{q}\lesssim10^{-3}\;.\label{kappaQval}%
\end{equation}
The first three coefficients scale linearly with $\lambda\lbrack
M_{\mathrm{GUT}}]$, so $\kappa_{1}^{q}[M_{\mathrm{SUSY}}]$ stays of the order
of $\lambda [M_{\mathrm{SUSY}}]$. The radiative feeding of
$\mathbf{A}_{udd}$ by $\mathbf{Y}_{udd}$ is thus quite intense; it is
impossible to have $\mathbf{A}_{udd}^{IJK}[M_{\mathrm{SUSY}}]\approx0$ once
$\lambda[ M_{\mathrm{GUT}}]$ is turned on. Also, note that even though
the leading coefficient is the largest, even a radiatively-induced
$\mathbf{A}_{udd}[M_{\mathrm{SUSY}}]$ is not holomorphic at the low scale. In
addition, these values are rather stable. If one starts with a non-vanishing
$\mathbf{A}_{udd}$ at the GUT scale, the RG evolution push the $\kappa_{i}%
^{q}$ coefficients back to the same values as in Eq.~(\ref{kappaQval}). As
shown in Fig.~\ref{kappaQ}, this fixed-point behavior is impressively
effective for the subleading coefficients. Thus, in the holomorphic case, not
only $\mathbf{Y}_{udd}[M_{\mathrm{SUSY}}]$ depends on the single input
parameter $\lambda\lbrack M_{\mathrm{GUT}}]$, but to a good approximation also
$\mathbf{A}_{udd}[M_{\mathrm{SUSY}}]$ since $\lambda\lbrack M_{\mathrm{GUT}}]$
defines the fixed-point values of Eq.~(\ref{kappaQval}).

\subsection{Holomorphy as an attractor}

If $\mathbf{Y}_{udd}$ is not holomorphic at some scale, it will remain so at all scales since the subleading expansion coefficients $\lambda_{i}^{q}$ of $\mathbf{Y}_{udd}^{Q}$ are non-zero. Looking back at Eq.~(\ref{LamdQ}), it is clear that these coefficients do not multiply RG invariant structures. Rather, through the evolution, each of these coefficients contribute a priori to all the others.

\begin{figure}[t]
\centering                  \includegraphics[width=15.5cm]{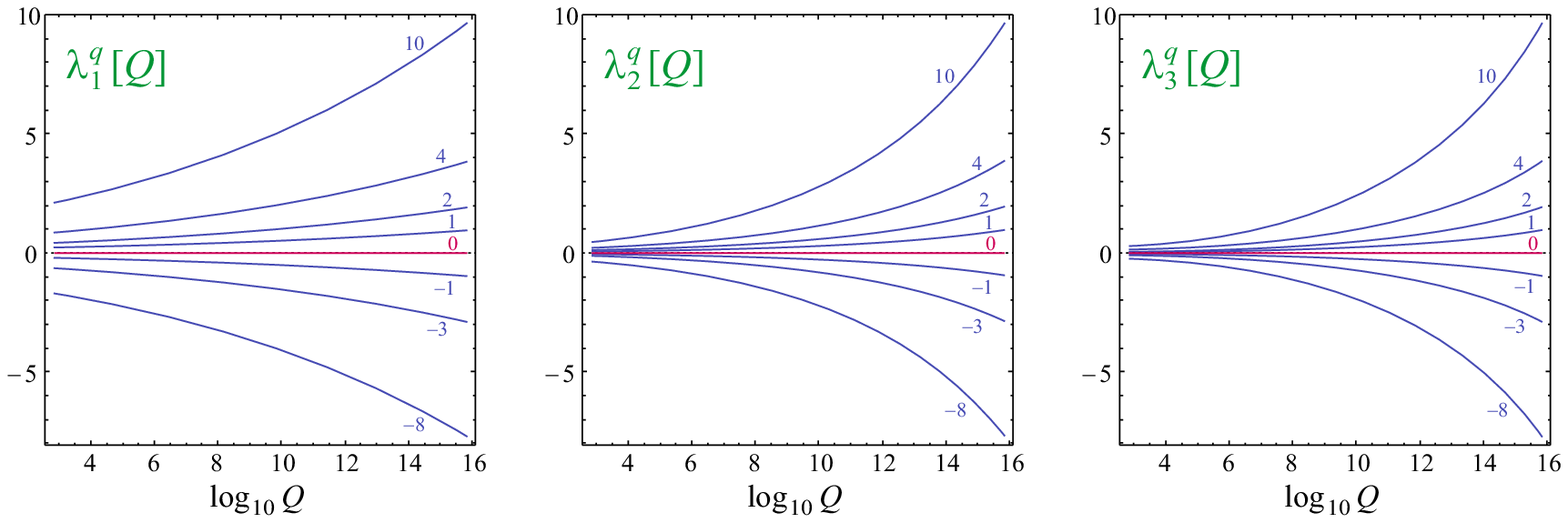}
\caption{Evolution of the leading coefficients of $\mathbf{Y}_{udd}$ in the broken $U(1)_{Q}$ scenario. The GUT scale boundary conditions are set as described in Appendix~\ref{CMSSM}, together with $\lambda_{i}^{q}[M_{\mathrm{GUT}}]=\lambda\delta_{i1}$ (left), $\lambda_{i}^{q}[M_{\mathrm{GUT}}]=\delta_{i1}+\lambda\delta_{i2}$ (middle), and $\lambda_{i}^{q}[M_{\mathrm{GUT}}]=\delta_{i1}+\lambda\delta_{i3}$ (right), with $\lambda=\{-8,-3,-1,0,1,2,4,10\}$. The left-hand plot corresponds to the pure holomorphic case, and shows the factor $\sim5$ reduction, see Eq.~(\ref{SolHolo}). The other two plots show that the subleading coefficients always converge towards zero, i.e., $\mathbf{Y}_{udd}$ runs towards a pure
holomorphic form.}%
\label{lambdaQ}%
\end{figure}

What is remarkable is that the holomorphic scenario of Ref.~\cite{CsakiGH11} emerges as an infrared (IR) fixed point. Specifically, starting from some non-zero $\lambda_{i\neq1}^{q}$ at the GUT scale, they all evolve towards much reduced values at the low scale. For example, starting with%
\begin{equation}
\lambda_{i}^{q}[M_{\mathrm{GUT}}]=(1,\;1,\;1,\;1,\;1,\;1,\;1,\;1,\;1)\;,
\end{equation}
we find that the leading coefficient is not affected by the others (we recover the same $\lambda_{1}^{q}$ value as in the previous section), while all the others are suppressed by more than an order of magnitude:%
\begin{align}
\lambda_{i}^{q}[M_{\mathrm{SUSY}}] &
=(0.221,\;0.049,\;0.031,\;0.079,\;0.023,\;0.025,\;0.013,\;-0.005,\;0.007)\nonumber\\
&
+i(0.000,\;0.000,\;0.000,\;0.000,\;-0.001,\;0.001,\;0.003,\;0.008,\;0.003)\;.
\end{align}
This convergence towards zero is effective even when the starting values $\lambda_{i\neq1}^{q}[M_{\mathrm{GUT}}]$ are much larger than one, as illustrated in Fig.~\ref{lambdaQ} for $\lambda_{2}^{q}$ and $\lambda_{3}^{q}$. The scaling between the values at the GUT and SUSY scale is mostly linear, with for example $\lambda_{i\neq1}^{q}[M_{\mathrm{GUT}}]\sim\mathcal{O}(100)$ leading to $\lambda_{i\neq1}^{q}[M_{\mathrm{SUSY}}]\sim\mathcal{O}(1)$. This observation has an important corollary: if any of the $\lambda_{i\neq1}^{q}$ is $\mathcal{O}(1)$ or larger at the low-scale, then they necessarily evolve towards non-MFV values at the GUT scale.

This behavior is similar to that of the coefficients of the RPC soft-breaking terms discussed in Refs.~\cite{Paradisi:2008qh,Colangelo:2008qp}, but for two differences. First, it is much more pronounced in the present case. The IR values are very small, $\lambda_{i\neq1}^{q}[M_{\mathrm{SUSY}}]\ll1$, while the RPC soft-breaking coefficients are $\mathcal{O}(1)$ at the low scale in general. Second, the values of the IR fixed points of all the $\lambda_{i\neq1}^{q}$ are trivially independent of the SUSY parameters since they are simply zero. On the contrary, for the RPC soft-breaking terms, the IR values depend on the MSSM parameters (gluino mass, scalar masses, etc), hence were dubbed ``quasi'' fixed points in Ref.~\cite{Colangelo:2008qp}.

The presence of this unique and true fixed point is of immediate phenomenological relevance\footnote{It should be stressed that the IR fixed points discussed here for the expansion coefficients have nothing in common with those discussed in Refs.~\cite{RPVRGE1, Ananthanarayan:1999bz} for the couplings themselves. In our case, the RPV couplings do not exhibit fixed points since the leading coefficient and the Yukawa couplings evolve in the IR.}. If MFV is active at some very high scale and if $U(1)_{Q}$ is broken, then to an excellent approximation, $\mathbf{Y}_{udd}[M_{\mathrm{SUSY}}]$ is holomorphic at the low scale since the subleading coefficients $\lambda_{i\neq1}^{q}[M_{\mathrm{SUSY}}]$ are tiny. The non-holomorphic corrections to $\mathbf{Y}_{udd}[M_{\mathrm{SUSY}}]$, which are in any case rather suppressed since they involve more Yukawa couplings, are thus entirely negligible and the whole baryonic RPV sector can be parametrized by a single parameter.

\subsection{Comparison with the broken $U(1)_{D}$ case}

\begin{figure}[t]
\centering                  \includegraphics[width=15.5cm]{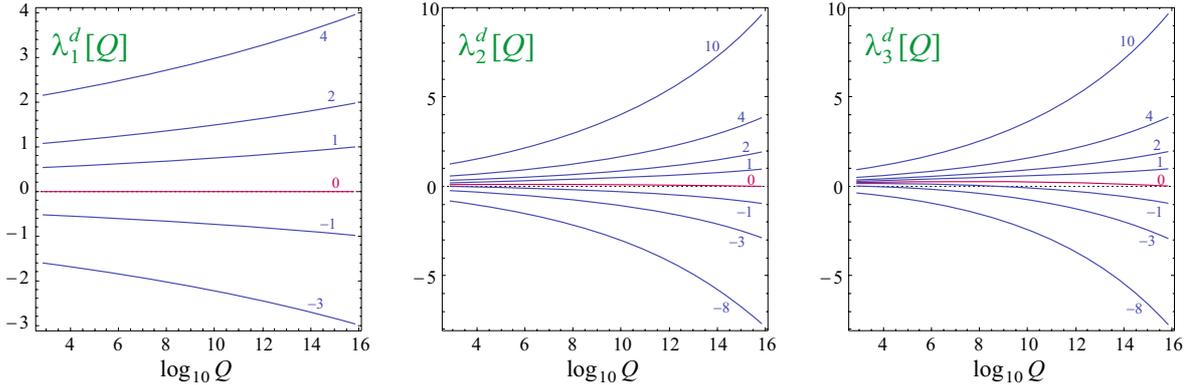}
\caption{Evolution of the leading coefficients of $\mathbf{Y}_{udd}$ in the broken $U(1)_{D}$ scenario. The GUT scale boundary conditions are set as described in Appendix~\ref{CMSSM}, together with $\lambda_{i}^{d}[M_{\mathrm{GUT}}]=\lambda\delta_{i1}$ (left), $\lambda_{i}^{d}[M_{\mathrm{GUT}}]=\delta_{i1}+\lambda\delta_{i2}$ (middle), and $\lambda_{i}^{d}[M_{\mathrm{GUT}}]=\delta_{i1}+\lambda\delta_{i3}$ (right), with $\lambda=\{-8,-3,-1,0,1,2,4,10\}$. For the left-hand plot, $|\lambda|$ is limited to be less than about $5$, otherwise the large $\mathbf{Y}_{tds}$ coupling renders the RGE numerically unstable. For the other two, the central curves (in red) correspond to the purely radiative generation, with the values at $M_{\mathrm{SUSY}}=1$~TeV given in  Eq.~(\ref{RunLamD}). }%
\label{lambdaD}%
\end{figure}

To illustrate how peculiar is the behavior of $\mathbf{Y}_{udd}^{Q}$, let us perform the same analysis starting with $\mathbf{Y}_{udd}^{D}$ instead. To start with, let us evolve down the leading $\mathbf{Y}_{udd}^{D}$ structure, i.e., set
\begin{equation}
\lambda_{i}^{d}[M_{\mathrm{GUT}}]=(1,\;0,\;0,\;0,\;0,\;0,\;0,\;0,\;0)\;.
\end{equation}
At the low-scale, the whole series of nine coefficients is generated:%
\begin{align}
\lambda_{i}^{d}[M_{\mathrm{SUSY}}]  &
=(0.54,\;0.10,\;0.25,\;0.06,\;0.09,\;0.02,\;0.03,\;0.04,\;0.00)\nonumber\\
&  +i(0.00,\;0.00,\;0.00,\;0.00,\;0.00,\;0.00,\;0.03,\;-0.03,\;0.03)\;.
\label{RunLamD}%
\end{align}
For comparison, if we now set all the $\mathbf{Y}_{udd}^{D}$ coefficients to one at the GUT scale
\begin{equation}
\lambda_{i}^{d}[M_{\mathrm{GUT}}]=(1,\;1,\;1,\;1,\;1,\;1,\;1,\;1,\;1)\;,
\end{equation}
we find the low-scale values%
\begin{align}
\lambda_{i}^{d}[M_{\mathrm{SUSY}}]  &
=(0.54,\;0.22,\;0.32,\;0.32,\;0.16,\;0.14,\;0.11,\;0.09,\;0.03)\nonumber\\
&  +i(0.00,\;0.00,\;0.00,\;0.00,\;0.01,\;0.00,\;0.01,\;-0.16,\;0.11)\;.
\end{align}

These examples show that MFV is preserved through the running, but the subleading coefficients are not particularly reduced at the low scale. Compared to the broken $U(1)_Q$ scenario, the leading coefficient $\lambda_{1}^{d}$ still evolves essentially independently of the others but the $\lambda_{i\neq1}^{d}$ do not converge towards zero. This can be seen in Fig.~\ref{lambdaD}, where the evolutions of $\lambda_{1}^{d}$, $\lambda_{2}^{d}$, and $\lambda_{3}^{d}$ are shown for various boundary conditions. Though a strong convergence of $\lambda_{2}^{d}$ and $\lambda_{3}^{d}$ towards their purely radiative values of Eq.~(\ref{RunLamD}) is apparent, these are not true fixed points. Indeed, being finite, they must necessarily depend on the specific MSSM scenario. In other words, for a different choice of MSSM parameters, $\lambda_{2}^{d}$ and $\lambda_{3}^{d}$ would run towards different values.

The existence of these IR fixed points implies that MFV at the low scale does not necessarily transcribe into MFV at the high scale. In view of Fig.~\ref{lambdaD}, it is clear that low scale values of a few units for $\lambda_{2}^{d}$ or $\lambda_{3}^{d}$ correspond to large non-MFV values at the GUT scale. So, imposing that MFV remains valid at all scales requires $\lambda_{i}^{d}[M_{\mathrm{SUSY}}]\lesssim1$ when $\lambda_{1}^{d}[M_{\mathrm{GUT}}]=1$. Note also that at the GUT scale, $\lambda_{1}^{d}[M_{\mathrm{GUT}}]$ cannot exceed a few units, because otherwise $\mathbf{Y}_{tds}^{D}[M_{\mathrm{GUT}}]>1$ and perturbativity collapses. In this respect, it must be stressed that when $\lambda_{1}^{d}[M_{\mathrm{GUT}}]\approx1$, the impact of $\mathbf{Y}_{udd}^{D}$ on the Yukawa couplings and on the soft-breaking terms is far from negligible. Given that the CMSSM parameters used throughout this work (see Appendix~\ref{CMSSM}) are quite fine-tuned to get a viable mass spectrum in the R-parity conserving case, especially a Higgs boson mass at around 125~GeV, the above numerical evaluations should be understood as illustrations for the behavior of the coefficients.

\begin{figure}[t]
\centering                  \includegraphics[width=15.5cm]{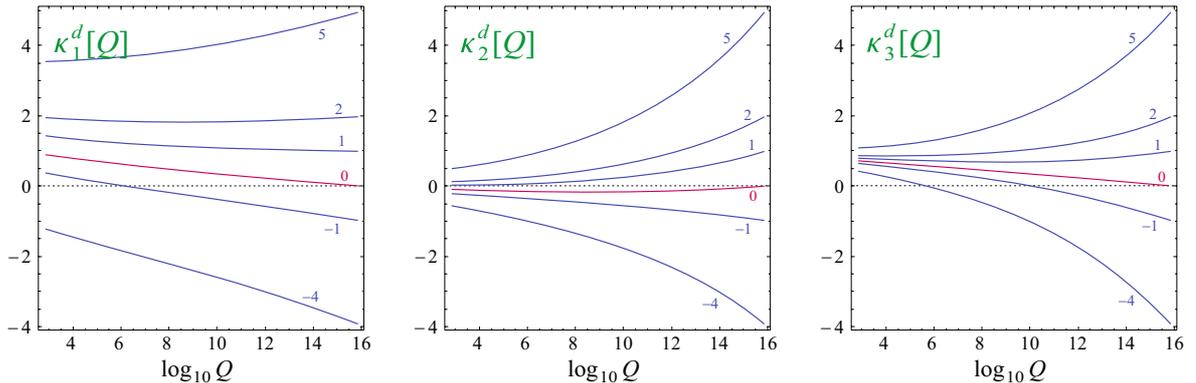}
\caption{Evolution of the leading coefficients for the trilinear couplings
$\mathbf{A}_{udd}$ in the broken $U(1)_{D}$ scenario. The GUT scale boundary
conditions are set as described in Appendix~\ref{CMSSM}, together with
$\lambda_{i}^{d}[M_{\mathrm{GUT}}]=\delta_{i1}$ and $\kappa_{i}^{d}%
[M_{\mathrm{GUT}}]=\kappa\delta_{ia}$ where $\kappa=\{-4,-1,0,1,2,5\}$ and
$a=1\ $(left), $a=2$ (middle), and $a=3$ (right). The central curves (in red)
correspond to the purely radiative generation, with the values at
$M_{\mathrm{SUSY}}=1$~TeV given in Eq.~(\ref{kappaDval}).}%
\label{kappaD}%
\end{figure}

Finally, in the soft-breaking sector, the radiative generation of
$\mathbf{A}_{udd}^{D}$ from $\mathbf{Y}_{udd}^{D}$ is even stronger than in
the holomorphic case, with for example%
\begin{align}
\kappa_{i}^{d}[M_{\mathrm{SUSY}}]  &
=(0.89,\;-0.12,\;0.69,\;0.02,\;0.23,\;-0.06,\;0.04,\;0.06,\;0.01)\nonumber\\
&  +i(0.00,\;0.00,\;0.00,\;0.00,\;0.00,\;0.00,\;0.05,\;-0.05,\;0.09)\;,
\label{kappaDval}%
\end{align}
when starting with $\kappa_{i}^{d}[M_{\mathrm{GUT}}]=0$ and $\lambda_{i}^{d}[M_{\mathrm{GUT}}]=\delta_{i1}$. If we instead allow for non-zero values for the $\mathbf{A}_{udd}^{D}[M_{\mathrm{GUT}}]$ coefficients, their low-scale values all respect the MFV principle, and a quasi-fixed point behavior is again apparent, see Fig.~\ref{kappaD}.

All in all, the evolution of the scenario based on $U(3)_{Q}\otimes U(3)_{U}\otimes SU(3)_{D}$ is less simple than for the one based on $SU(3)_{Q}\otimes U(3)_{U}\otimes U(3)_{D}$, which is very peculiar because of the presence of the holomorphic true fixed point. Still, the behaviors of the coefficients of $\mathbf{Y}_{udd}^{D}$ and $\mathbf{A}_{udd}^{D}$ remain rather smooth, MFV is preserved down from the GUT scale, and quasi fixed points in the IR are apparent (the same could be said for $\mathbf{Y}_{udd}^{U}$ and $\mathbf{A}_{udd}^{U}$ in the broken $U(1)_U$ scenario). Actually, this is perfectly in line with the behaviors of the coefficients of the RPC sector~\cite{Paradisi:2008qh,Colangelo:2008qp}. Therefore, in view of the large $\mathbf{Y}_{tds}^{D}$ coupling, this scenario is worth investigating further, both experimentally and phenomenologically. In particular, a dedicated study of the impact of a $\mathbf{Y}_{tds}^{D}%
\sim\mathcal{O}(1)$ coupling on the supersymmetric spectrum, and thus indirectly on the Higgs boson mass, could prove very valuable, not least because it could in principle lessen the puzzling fine-tunings, and enlarge the viable MSSM parameter space.

\section{Conclusions}

In this paper, we have analyzed the behavior of the R-parity violating couplings under the renormalization group evolution. Particular emphasis is laid on the MFV restriction, since it permits to naturally pass all the bounds from proton decay or neutron oscillations even for relatively light superparticles~\cite{RPVMFV}. To this end, the formulation of the MFV hypothesis in the RPV sector first had to be made more precise and robust. Specifically, our procedure and our main results are:

\begin{enumerate}
\item We have constructed a basis-independent parametrization for completely generic baryonic R-parity violating couplings, in a way similar to that proposed for the R-parity conserving soft-breaking terms of the slepton sector in Ref.~\cite{Mercolli:2009ns} and of the squark sector in Ref.~\cite{NatSUSY}. It trades the 18 independent RPV couplings for 18 free expansion parameters, whose numerical values are independent of the flavor basis chosen for the (s)quark fields. As such, they thus fully encode the RPV sector. For example, experimental constraints translate into bounds on these coefficients, MFV is directly obtained by restricting these coefficients to $\mathcal{O}(1)$ values, they permit to set in an unambiguous way any boundary conditions for the RPV couplings, and they even suffice to describe the whole RG evolution of the RPV couplings.

\item We have shown that to impose MFV on the whole R-parity violating MSSM and at vastly different scales, it is necessary to restrict the flavor symmetry group. Out of the $U(3)^{3}$ symmetry of the (s)quark kinetic terms, only one $U(1)$ can be broken at a time. Failure to do so generates ambiguities in the phases of the expansion coefficients, renders the $\mathcal{O}(1)$ naturality criterion for expansion coefficients ambiguous, and even invalidates the usual MFV expansions in the R-parity conserving sector. On the other hand, once properly set, the multi-scale MFV hypothesis is rather resilient. The RG evolution of the RPV expansion coefficients displays striking infrared fixed or quasi-fixed point behavior, ensuring that MFV at the low scale arises even from far from-MFV scenarios at the unification scale. The corollary also holds: if the expansion coefficients at the low scale are $\mathcal{O}(1)$ but far from their (quasi-) fixed points, then MFV is lost at the unification scale. In these respects, the RPV sector behaves very similarly to the R-parity conserving soft-breaking sector~\cite{Paradisi:2008qh,Colangelo:2008qp}.

\item Finally, we have explored the RG behavior of the holomorphic MFV scenario. First, we proved analytically that the holomorphic restriction is RG invariant. This is far from trivial since the Yukawa spurions are never true dynamical fields, but has far-reaching consequences. In particular, it implies that holomorphy acts as a powerful infrared attractor for the RG evolution. If present at the high scale, all the non-holomorphic corrections evolve towards zero at the low scale. Whether exact or approximate, low-scale holomorphy thus systematically emerges as the phenomenological paradigm once the broken flavor $U(1)$ is that of the quark doublet. Intriguingly, this same flavored $U(1)$ is the one already broken by the $\mathcal{B}+\mathcal{L}$ anomaly of the weak interactions~\cite{tHooft176,tHooft276}. In other words, MFV, which must obviously hold in the SM, is compatible with the $\mathcal{B}+\mathcal{L}$ anomaly only if that $U(1)$ is broken~\cite{MFVBL}. Though the connection appears rather coincidental at present, it is thus tempting to conclude that low-scale holomorphy should hold, at least to a good approximation.
\end{enumerate}

Imposing MFV on the R-parity violating couplings is not only a viable phenomenological approach, but it may also hint at some more fundamental aspects of the yet unknown origin to the flavor structures. For instance, the allowed flavor symmetry group and the RG properties of their MFV implementations may prove crucial in the search for a dynamical realization of MFV at some very high scale. Besides this model-building frontier, what is still lacking to fully validate this approach is, obviously, some experimental signals. We are thus eagerly waiting for the next round of searches at the LHC, which will hopefully usher us beyond the realm of the Standard Model.

\subsubsection*{Acknowledgments}

We would like to thank Sabine Kraml for discussions and comments. This work has been supported in part by the ``Investissements d'avenir, Labex ENIGMASS''.

\pagebreak 

\appendix                                                           

\section{Cayley-Hamilton theorem and matrix identities\label{CHth}}

Combining the definition of the determinant, $\varepsilon^{LMN}A^{LI}%
A^{MJ}A^{NK}\equiv\det(A)\varepsilon^{IJK}$, with the Cayley-Hamilton theorem%
\begin{equation}
A^{3}-\left\langle A\right\rangle A^{2}+\tfrac{1}{2}A(\langle A\rangle
^{2}-\langle A^{2}\rangle)=\det A\;,\label{CH}%
\end{equation}
where $A$ is a generic three-by-three complex matrix, leads to several useful
identities. The starting point is to take the trace of the identity~(\ref{CH})
to express $\det A$ in terms of traces of powers of $A$ only:%
\begin{equation}
\varepsilon^{LMN}A^{LI}A^{MJ}A^{NK}\equiv\det(A)\varepsilon^{IJK}%
=\varepsilon^{IJK}[\;\tfrac{1}{3}\langle A^{3}\rangle-\tfrac{1}{2}\langle
A\rangle\langle A^{2}\rangle+\tfrac{1}{6}\langle A\rangle^{3}%
\;]\;.\label{DetCH}%
\end{equation}
From this, we can derive simpler identities involving traces and antisymmetric
contraction by shifting $A\rightarrow1+A$, expand in $A$, and extract terms
linear and quadratic in $A$:%
\begin{subequations}
\begin{align}
\varepsilon^{LJK}A^{LI}+\varepsilon^{ILK}A^{LJ}+\varepsilon^{IJL}A^{LK} &
=\varepsilon^{IJK}\langle A\rangle\;,\label{Det1CHa}\\
\varepsilon^{LMK}A^{LI}A^{MJ}+\varepsilon^{LJM}A^{LI}A^{MK}+\varepsilon
^{ILM}A^{LJ}A^{MK} &  =\varepsilon^{IJK}\tfrac{1}{2}[\,\langle A\rangle
^{2}-\langle A^{2}\rangle\,]\;.
\end{align}
Other useful identities are derived by multiplying the definition of the
determinant by $A^{-1}$,%
\end{subequations}
\begin{equation}
(A^{-1})^{PK}\det(A)\varepsilon^{IJP}=\varepsilon^{LMN}A^{LI}A^{MJ}%
A^{NP}(A^{-1})^{PK}=\varepsilon^{LMK}A^{LI}A^{MJ}\;.
\end{equation}
The left-hand side can be simplified using the Cayley-Hamilton theorem.
Multiplying both sides of Eq.~(\ref{CH}) by $A^{-1}$ leads to
\begin{equation}
\varepsilon^{ILM}A^{LJ}A^{MK}=\varepsilon^{LJK}[\,A^{2}-\langle A\rangle
A+\tfrac{1}{2}\langle A\rangle^{2}-\tfrac{1}{2}\langle A^{2}\rangle
\,]^{LI}\;.\label{Det2CHb}%
\end{equation}
Finally, there are also identities with several different matrices. For
example, by plugging $A\rightarrow A+B$ in Eq.~(\ref{Det2CHb}), we can derive
\begin{equation}
\varepsilon^{ILM}(A^{LJ}B^{MK}+B^{LJ}A^{MK})=\varepsilon^{LJK}\left[
\,\{A,B\}-\langle A\rangle B-\langle B\rangle A+\langle A\rangle\langle
B\rangle-\langle AB\rangle\,\right]  ^{LI}\;.
\end{equation}
The most general three-matrix identity is found by replacing $A\rightarrow
A+B+C$ in Eq.~(\ref{DetCH}),%
\begin{equation}
\varepsilon^{LMN}\{A,B,C\}^{LI,MJ,NK}=\varepsilon^{IJK}\left[  \langle
ABC+ACB\rangle-\langle A\rangle\langle CB\rangle-\langle B\rangle\langle
AC\rangle-\langle C\rangle\langle AB\rangle+\langle A\rangle\langle
B\rangle\langle C\rangle\right]  ,\label{GenID}%
\end{equation}
where $\{A,B,C\}^{a,b,c}\equiv A^{a}B^{b}C^{c}+A^{a}C^{b}B^{c}+B^{a}A^{b}%
C^{c}+B^{a}C^{b}A^{c}+C^{a}A^{b}B^{c}+C^{a}B^{b}A^{c}$. From this, simpler
identities can be obtained by setting some matrices to $\mathbf{1}$ and/or
equating some of them. For example, when $C=A$, Eq.~(\ref{GenID}) reduces to
\begin{multline}
\;\;\;\;\;\;\;\;\;\;\;\;\varepsilon^{LMN}(A^{LI}A^{MJ}B^{NK}+A^{LI}%
B^{MJ}A^{NK}+B^{LI}A^{MJ}A^{NK})\\
=\varepsilon^{IJK}\left[  \,\langle A^{2}B\rangle-\langle A\rangle\langle
BA\rangle+\tfrac{1}{2}\langle B\rangle(\langle A\rangle^{2}-\langle
A^{2}\rangle)\,\right]  \;.\;\;\;\;\;\;\;\;\;\;\;\;
\end{multline}

\section{Numerical procedures and mass spectrum\label{CMSSM}}

Throughout this work, we illustrate the behavior of the RPV couplings using a CMSSM-like scenario with $\tan\beta=10$ at the low scale. At the GUT scale, the boundary conditions are set as $m_{1/2}=m_{0}=-A_{0}/2=1~$TeV, $m_{H_{u}}^{2}=m_{H_{d}}^{2}=(1.2m_{0})^{2}$ and $\mathbf{m}_{Q,U,D,L,E}^{2}=m_{0}^{2}\mathbf{1}$, $\mathbf{A}_{u,d,e}=A_{0}\mathbf{Y}_{u,d,e}$. We do not fully unify the scalar masses in order to reproduce the observed Higgs boson mass. Specifically, in the RPC case, running these values through \texttt{SPheno 3.2.4}~\cite{SPheno1,SPheno2}, the Higgs sector mass spectrum is $m_{h^{0}}=123~$GeV together with $m_{A^{0}}\approx m_{H^{0}}\approx m_{H^{\pm}}=2.0~$TeV, while the SUSY spectrum is%
\begin{align}
m_{\tilde{g}}  &  =2.2~\text{TeV\ ,\ \ }m_{\tilde{\chi}^{\pm}}%
=(0.82,\;1.5)~\text{TeV\ ,\ \ }m_{\tilde{\chi}^{0}}%
=(0.43,\;0.82,\;1.5,\;1.5)~\text{TeV\ ,}\\
m_{\tilde{u}}  &  =(1.4,\;1.9,\;2.2,\;2.2,\;2.2,\;2.2)~\text{TeV\ ,\ }%
m_{\tilde{d}}=(1.9,\;2.1,\;2.1,\;2.1,\;2.2,\;2.2)~\text{TeV\ ,}\\
m_{\tilde{e}}  &  =(1.0,\;1.1,\;1.1,\;1.2,\;1.2,\;1.2)~\text{TeV\ ,\ }%
m_{\tilde{\nu}}=(1.2,\;1.2,\;1.2)~\text{TeV\ .}%
\end{align}
At the low scale, the expansion coefficients for the RPC soft-breaking terms respect the MFV principle, with for example (see Eq.~(\ref{RPCSoft}))%
\begin{equation}
a_{i}^{q}=(4.8,\;-3.0,\;-2.4,\;0.27,\;0.87,\;0.93,\;0.003,\;-0.004,\;-0.004)\;,
\end{equation}
This combined with the heavy sparticle masses ensure that all the flavor observables are in check in this scenario.

Turning on the RPV couplings, the numerical evolutions are not computed with \texttt{SPheno}. Rather, we use custom \texttt{Mathematica} programs to solve the one-loop RG equations~\cite{RPVRGE1, RPVRGE2} of the RPV-MSSM between $M_{\mathrm{SUSY}}$ and $M_{\mathrm{GUT}}$. We check that they agree at the percent level with \texttt{SPheno} in the CP- and R-parity conserving case.

The RPV couplings $\mathbf{Y}_{udd}$ and $\mathbf{A}_{udd}$ are set at the GUT
scale in a basis-independent way through their expansion coefficients. The
multiscale boundary conditions $\mathbf{Y}_{u,d,e}[M_{\mathrm{SUSY}}]$ and
$\mathbf{Y}_{udd}[M_{\mathrm{GUT}}]$, with in addition $\mathbf{Y}%
_{udd}[M_{\mathrm{GUT}}]=F(\mathbf{Y}_{u,d,e}[M_{\mathrm{GUT}}])$ for some
expansion $F$, are imposed iteratively. Starting with $\mathbf{Y}_{udd}%
^{0}[M_{\mathrm{SUSY}}]=F(\mathbf{Y}_{u,d,e}[M_{\mathrm{SUSY}}])$, a few
iterations permit to find an input value $\mathbf{Y}_{udd}[M_{\mathrm{SUSY}}]$
such that it evolves up to $F(\mathbf{Y}_{u,d,e}[M_{\mathrm{GUT}}])$. This is
rather fast since only the supersymmetric parameters are involved at that stage.

We do not derive the RG equations for the expansion coefficients of
$\mathbf{Y}_{udd}$ (nor of $\mathbf{A}_{udd}$). Instead, their evolutions are
obtained indirectly by projecting $\mathbf{Y}_{udd}[M_{Q}]$ on one of the
bases written in terms of $\mathbf{Y}_{u,d}[M_{Q}]$ at various intermediate
scales $M_{Q}$. This means each time solving a linear system of nine
equations, one for each of the independent entries of $\mathbf{Y}_{udd}%
[M_{Q}]$. In practice, this is trickier than it seems because of the very
large hierarchies involved. For example, the largest and smallest couplings of
$\varepsilon^{LMN}(\mathbf{Y}_{u}\mathbf{X}_{u}^{2})^{IL}\mathbf{Y}_{d}%
^{JM}\mathbf{Y}_{d}^{KN}$ are 22 orders of magnitude apart (because of a
factor $m_{u}^{4}/m_{t}^{4}$). To avoid spuriously large coefficients when the
system of equations is solved exactly, both the evolution and the matching at
$M_{\mathrm{GUT}}$ have to be done using more than the default 16 digits of
precision. Alternatively, if one sticks to the default precision, one should
approximately solve the system of equations, under the constraint that the
coefficients are the smallest as possible. We have not implemented that
solution, because forcing \texttt{Mathematica} to work with somewhere between
25 to 30 digits does not prohibitively slow down the various routines.

We do not include the RPV threshold corrections, nor do we study the
corrections to the Higgs mass brought in by $\mathbf{Y}_{udd}$, since our goal
here is to illustrate the evolution of the RPV couplings. This is a very good
approximation when the RPV couplings respect the MFV principle with either
$U(1)_{Q}$ or $U(1)_{U}$ broken, since $\mathbf{Y}_{udd}$ is then very
suppressed, see Table~\ref{Hierarchies}. On the other hand, when $U(1)_{D}$ is
broken, the $\mathbf{Y}_{tds}$ coupling is large and can in principle affect
significantly the spectrum. Though we neglect this effect here, it should be
kept in mind.

Finally, let us stress that this benchmark is not tailored to induce an
interesting phenomenology at the LHC. With its rather heavy spectrum, most
signatures are suppressed, like for example in the same-sign top quark pair
channel discussed e.g. in
Refs.~\cite{Allanach:2012vj,AsanoRS13,DurieuxEtAl,BergerPST13,DurieuxS13,Batell:2013zwa,Saelim:2013gea,Allanach:2013qna,Bhattacherjee:2013tha}%
. Indeed, though the neutralino LSP would decay exclusively to top quarks when
either $U(1)_{Q}$ or $U(1)_{D}$ is broken, its pair production is not intense
when all squark masses are above the TeV scale. To find more interesting
benchmarks for the LHC is certainly interesting, but beyond our scope here.

\bibliography{biblio}
\bibliographystyle{utphys}

\end{document}